\documentclass[twocolumn,prb,a4,showpacs,superscriptaddress,reprint,footinbib]{revtex4-1}
\usepackage[]{graphicx}
\usepackage{epstopdf,braket,amsmath,amssymb,hyperref,comment,color,relsize}
\usepackage[T1]{fontenc}
\usepackage[utf8]{inputenc}
\definecolor{myOrange}{RGB}{255,128,0}
\pdfoutput=1

\begin{document}
	\title{Ballistic electron channels including weakly protected topological states in delaminated bilayer graphene.}
	\author{T.L.M. Lane}
	\email{thomas.lane-3@postgrad.manchester.ac.uk}
	\affiliation{National Graphene Institute, University of Manchester, Manchester, M13 9PL, UK}
	\affiliation{School of Physics and Astronomy, University of Manchester, Manchester, M13 9PL, UK}
	\author{M. An\dj{}elkovi\'c}
	\affiliation{Department Fysica, Universiteit Antwerpen, Groenenborgerlaan 171, B-2020 Antwerpen, Belgium}
	\author{J.R. Wallbank}
	\affiliation{National Graphene Institute, University of Manchester, Manchester, M13 9PL, UK}
	\author{L. Covaci}
	\affiliation{Department Fysica, Universiteit Antwerpen, Groenenborgerlaan 171, B-2020 Antwerpen, Belgium}
	\author{F.M. Peeters}
	\affiliation{Department Fysica, Universiteit Antwerpen, Groenenborgerlaan 171, B-2020 Antwerpen, Belgium}
	\affiliation{National Graphene Institute, University of Manchester, Manchester, M13 9PL, UK}
	\author{V.I. Fal'ko}
	\affiliation{National Graphene Institute, University of Manchester, Manchester, M13 9PL, UK}
	\affiliation{School of Physics and Astronomy, University of Manchester, Manchester, M13 9PL, UK}
	
	\begin{abstract}
		We show that delaminations in bilayer graphene (BLG) with electrostatically induced interlayer asymmetry can provide one with ballistic channels for electrons with energies inside the electrostatically induced BLG gap. These channels are formed by a combination of valley-polarised evanescent states propagating along the delamination edges (which persist in the presence of a strong magnetic field) and standing waves bouncing between them inside the delaminated region (in a strong magnetic field, these transform into Landau levels in the monolayers). For inverted stacking between BLGs on the left and right of the delamination (AB-2ML-BA or BA-2ML-AB), the lowest energy ballistic channels are gapless, have linear dispersion and appear to be weakly topologically protected. When BLG stacking order on both sides of the delamination is the same (AB-2ML-AB or BA-2ML-BA), the lowest energy ballistic channels are gapped, with gap $\varepsilon_g$ scaling as $\varepsilon_g\propto W^{-1}$ with delamination width and as $\varepsilon_g\propto\delta^{-1}$ with the on-layer energy difference within the delamination. Depending on their width, delaminations may also support several `higher energy' waveguide modes. Our results are based on both an analytical study of the wavematching of Dirac states and tight binding model calculations, and we analyse in detail the dependence of the delamination spectrum on electrostatic conditions in the structure, such as the vertical displacement field.
	\end{abstract}
	
	\maketitle
	
	
	\section{Introduction}
	
	Demand for increasingly dense computational architectures is driving the miniaturisation of conventional electronic circuits to their limit, requiring novel technologies to be developed. Single layer graphene, with its gapless band structure, high mobility carriers, and high thermal conductivity \cite{novoselov_nature_2005} has been considered as a candidate for the creation of conducting nano-channels. However, lithographic processes used for the patterning of such wires spoil graphene edges, introducing defects which make fabrication of ballistic channels in graphene a technological challenge. At the same time, the use of a split-gated structure on monolayer graphene does not help to confine electrons due to high transparency of p-n interfaces \cite{cheianov_PRB_2006}.
	
	An alternative approach to creating ballistic `one-dimensional' channels in graphene is to use a gate-controlled gap in its Bernal (AB) stacked bilayer allotrope, which has an electrostatically tunable band gap \cite{mccann_PRL_2006,mccann_PRB_2006,castro_PRL_2007,zou_PRB_2010,mucha_semiSci_2010}. Earlier studies \cite{martin_PRL_2008,zarenia_PRB_2011,daCosta_PRB_2015,cosma_PRB_2015,li_natNano_2016} have shown that sharply switching the direction of the vertical displacement field across these split-gated structures leads to ballistic `topological' modes localised near the boundaries, persisting across junctions of interfaces \cite{qiao_nanoLett_2011,wang_PRB_2017}. It has also been found that delaminations of bilayer graphene (BLG) can provide well-defined one-dimensional channels \cite{pelc_PRB_2015,abdullah_EPL_2016} with counter-propagating modes in the opposite valleys ($K^+$ and $K^-$). Experimental observations of states along delaminations naturally occurring in mechanically exfoliated graphene have already been reported \cite{ju_natLett_2015,yin_natComms_2016}.
	
	\begin{figure}[htb]
		\includegraphics[width=0.48\textwidth]{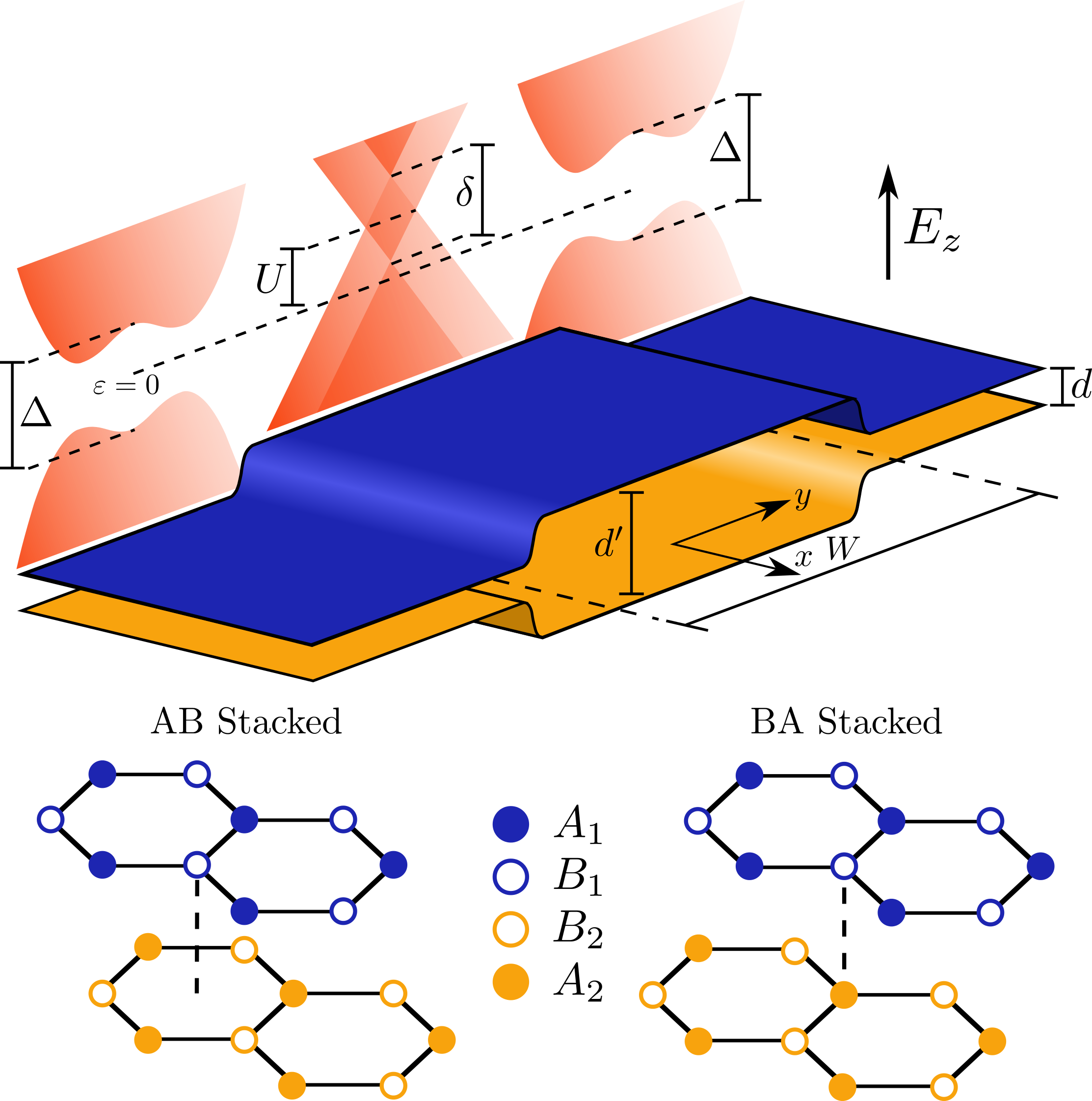}
		\caption{Sketch of the considered delamination. The decoupled monolayers lie in region $0<y<W$, connected to two bilayer graphene regions ($y\leq0$ \& $y\geq W$). Corresponding energy dispersions within each distinct zone are shown with the electrostatic variables of the model indicated. These would be tuned by varying the vertical displacement field, $E_z$, across the system via top and bottom gates in each region. Below are images depicting the two different Bernal stacking arrangements discussed in the text.}
		\label{fig01:delaminationSketch}
	\end{figure}
	
	In this paper, we study electronic properties of delaminations in a BLG sheet and their dependence on the electrostatically controlled displacement field applied to the structure. Such a system, modelled by both 4-band $\vec{k}\cdot\vec{p}$ theory of BLG and the numerical tight-binding (TB) model approach, is illustrated in Fig.~\ref{fig01:delaminationSketch}. The delamination is considered as two de-coupled monolayers of graphene (2ML) between two BLG regions in either the same (AB-2ML-AB or BA-2ML-BA) stacking configuration, or with opposite (AB-2ML-BA or BA-2ML-AB) stacking. The vertical displacement field $E_z$ (in real devices, controlled by top and bottom gates) induces a band gap in the BLG, $\Delta=eE_zd$, and also mutually shifts on-site energies on the two delaminated monolayers by $\delta=eE_zd'$.

	The low-energy band structure of Bernal stacked (AB or BA) BLG is described by a 4-band $\vec{k}\cdot\vec{p}$ Hamiltonian \cite{mccann_PRB_2006,mucha_semiSci_2010} and band dispersion,
	\begin{eqnarray}
	&&E^{BLG} =\label{BLG_eigVal}\\ &&\pm\sqrt{v^2(k_x^2-k_y^2)+\frac{\Delta^2}{4}+\frac{\gamma_1^2}{2}+r\sqrt{\frac{\gamma_1^4}{4}+v^2(k_x^2+k_y^2)(\gamma_1^2+\Delta^2)}}.\nonumber
	\end{eqnarray}
	Here, $r=\pm$ indexes the low and high-energy BLG bands, $\Delta$ is the interlayer asymmetry gap, $k_{x,y}$ are the in-plane wavevectors of electrons in $K^\pm$ valleys, and $\gamma_1=0.39$~eV and $v=6.6$~eV\AA\ are the interlayer coupling and Dirac velocity respectively \cite{neto_revModPhys_2009}. This BLG spectrum suggests that at the energies $|\varepsilon|<\varepsilon_*$,
	\begin{equation}
	\varepsilon_*=\frac{\gamma_1|\Delta|}{2\sqrt{\Delta^2+\gamma_1^2}},\label{BLG_lim}
	\end{equation}
	there are no states for electrons in a gapped bilayer.
	
	In contrast, the electron spectrum in the delaminated region is gapless (see Fig.~\ref{fig01:delaminationSketch}),
	\begin{equation}
	E^{2ML} = U+l\frac{\delta}{2}\pm v\sqrt{k_x^2-k_y^2}.\label{2ML_eigVal}
	\end{equation}
	Here $U$ is an energy shift between the BLG and 2ML regions, $l=\pm$ is upper(+)/lower(-) monolayer index, and $\delta$ is the energy offset between them produced by the displacement field (in principle, $\delta=eE_zd'$ can be larger than the BLG gap, $\Delta$, due to the larger interlayer distance within the delamination, $d'$>$d$). Due to its continuous spectrum, the delamination can support states within the BLG gap, and, in the following, we will analyse the dispersion of electrons channelled by the delamination.
	
	In particular, in Section \ref{singleIntSec} we analyse a system with a single interface between bias-gapped BLG and 2ML, which supports valley-polarised evanescent modes with linear dispersion.  In Section \ref{doubleIntSec}, we study evanescent edge states and standing waves inside a delamination using both the continuum $\vec{k}\cdot\vec{p}$ model and tight binding (TB) calculations. The form of these states depends on the choice of interlayer stacking on either side of the decoupled monolayers. Inequivalent BLG stacking on either side of the delamination (AB on one side and BA on the other) results in the valley-polarised channels propagating in the same direction along both interfaces, leading to a gapless dispersion with two weakly topologically protected modes with dispersion spanning across the BLG gap. Having the same stacking in the outer BLG parts of the structure reverses the direction of one of these channels, so that the resulting counter-propagating evanescent modes hybridise, producing a gapped spectrum. We also analyse the higher-energy `gapped' modes resulting from standing waves bouncing between the gapped BLG regions, and we study the dependence of the spectrum on the displacement-field-shifted energies of Dirac points in the delaminated layers. In Section \ref{magFieldSec}, we study how a strong magnetic field transforms the modes in the AB-2ML-BA structure into Landau levels in the delaminated monolayers.
	
	Before going into technical details of Sections \ref{singleIntSec}-\ref{magFieldSec}, we note that edges in graphene flakes can take two different forms; zig-zag and armchair \cite{akhmerov_PRB_2008,nakanishi_PRB_2010}. For the armchair edge, electron scattering from it mixes electron states in the two valleys. In contrast, for a zig-zag edge or arbitrarily cut edge, the large momentum difference between $K^+$ and $K^-$ projections onto the delamination axis suppresses intervalley mixing \cite{akhmerov_PRB_2008}. For this study, we assume general boundary conditions that coincide with those of a zig-zag edge, but will underline features of the armchair edge in Discussions Section \ref{discussionsSec}.

	\section{Electronic Properties of a Single BLG-2ML Interface}\label{singleIntSec}
	
	\begin{subequations}
	To study a single BLG-2ML interface using $\vec{k}\cdot\vec{p}$ theory, we employ a 4-band Hamiltonian,
	\begin{equation}
		\hat{{\cal H}}_\eta=\left(\begin{array}{c c c c}
			V_+(y) & v\hat{\pi}^\dagger & \Theta(-y)\gamma_1 & 0\\
			v\hat{\pi} & V_+(y) & 0 & w_3^\dagger\\
			\Theta(-y)\gamma_1 & 0 & V_-(y) & v\hat{\pi}\\
			0 & w_3 & v\hat{\pi}^\dagger & V_-(y)
		\end{array}\right),\label{singleInt_H}
	\end{equation}	
	written in the sub-lattice basis $(A_1,B_1,B_2,A_2)^T$ in valley $K^\eta$, where $\eta=\pm$, $\hat{\pi}=\eta(-i\partial_x)+i(-i\partial_y)$, and $\left\{\Theta(-y),\hat{\pi}\right\}$ in $w_3=\frac{v\gamma_3}{2\gamma_0}e^{i\theta}\left\{\Theta(-y),\hat{\pi}\right\}$ is an anti-commutator. In this Hamiltonian, hopping parameter $\gamma_1$ describes coupling of the `dimer' sites, $A_1$ and $B_2$, of the bilayer, whilst terms with $\gamma_3$ describe skew hopping (between non-dimer sites)\cite{gamma3Footnote}, and angle $\theta$ is between the zig-zag direction and delamination axis. On-site energies, $V_\pm(y)$, are defined as,
	\begin{equation}
		V_\pm(y)=
		\begin{cases}
			\pm\frac{\Delta}{2} &,  y\leq0 \\
			U\pm\frac{\delta}{2} &,  y>0
		\end{cases},\nonumber
	\end{equation}
	and $\Theta(y)$ is the Heaviside step function used to implement the suppression of $\gamma_1$ by delamination. This Hamiltonian has the dispersion of Eq.~(\ref{BLG_eigVal}) and Eq.~(\ref{2ML_eigVal}) in regions $y\leq0$ and $y>0$ respectively. Implicit within this Hamiltonian is the required continuity of the electron wavefunctions at the junction between the two regions,
	\begin{equation}
	\left(\begin{array}{c c c c}
	1 & 0 & 0 & -\aleph \\
	0 & 1 & 0 & 0 \\
	0 & -\aleph & 1 & 0 \\
	0 & 0 & 0 & 1
	\end{array}\right)\psi(y\rightarrow 0^-)=\psi(y\rightarrow 0^+)\label{matching}.
	\end{equation}
	\end{subequations}
	Parameter $\aleph=e^{i\theta}\gamma_3/(2\gamma_0)$ is determined by the ratio of the skew interlayer hopping \cite{mccann_PRL_2006} and intralayer hopping. As $\gamma_3/\gamma_0\approx0.12$ (hence, $\aleph\ll1$), the skew hopping ($\gamma_3$) terms lead to only small corrections to the interface mode dispersion and wavefunctions (see Fig.\ref{fig03:singleIntModes}). Therefore, the terms describing skew hopping will be neglected in the rest of the text (formally, setting $\gamma_3/\gamma_0\rightarrow0$).
	
	\subsection{Reflection at the BLG-2ML interface}
	
	Here we investigate the reflection of plane waves in a semi-infinite delamination and their scattering between two monolayers, from a gapped BLG. On the delaminated side of the interface we solve Dirac equation,
	\begin{equation}
	(\hat{{\cal H}}_\eta-\varepsilon\hat{I})\psi=0,\nonumber
	\end{equation}
	assuming propagating plane-wave solutions,
	\begin{eqnarray}
	&&\Psi^{2ML}=\psi^{2ML}e^{ik_xx}\propto e^{i(q_{s,l})y}e^{ik_xx},\nonumber\\
	&&q_{s,l}=s\frac{1}{v}\sqrt{\left(\varepsilon-U-l\frac{\delta}{2}\right)^2-v^2k_x^2},\label{2ML_lambda}
	\end{eqnarray}
	where $q_{s,l}$ would have real values. Here, $s=\pm1$ and $l=+(-)1$ distinguish between right/left moving waves ($y>0$) and the upper (lower) layers respectively:
	\begin{eqnarray}
	&&\psi^{2ML}(k_x,\varepsilon)=\nonumber\\[0.25\baselineskip]
	&&\sum_{s}\left[
	A_{s,+}\left(\begin{array}{c}
	1 \\ a_{s,+} \\ 0 \\ 0
	\end{array}\right)e^{iq_{s,+}y}+
	A_{s,-}\left(\begin{array}{c}
	0 \\ 0 \\ 1 \\ a_{s,-}
	\end{array}\right)e^{iq_{s,-}y}\right],\label{2ML_state}
	\end{eqnarray}
	where $A_{s,\pm}$ are the monolayer wave amplitudes, and
	\begin{equation}
	a_{s,l}=\frac{2v[k_x+ilq_{s,l}]}{2\varepsilon-2U-l\delta}
	\end{equation}
	are chirality factors for electrons in monolayer graphene.
	
	Because the direction of electron's propagation is given by group velocity, $\bar{v}=\partial\varepsilon/\partial k$, electrons in the monolayer conduction band with wavevector $\vec{k}$ move in the opposite direction to those in the monolayer valence band with the same wavevector. In order to distinguish between left and right moving states we write index $s=\pm\xi_l$, where $\xi_l=\text{sign}\left(\varepsilon-[U+l~\delta/2]\right)$ determines whether electrons lie in the conduction ($\xi_l=+$) or valence ($\xi_l=-$) band in each monolayer and the $\pm$ selects left ($-$) or right ($+$) moving states.
	
	At the same time, the asymptotics of eigenstates in the gapped bilayer region must be decaying,
	\begin{eqnarray}
	&&\Psi^{BLG}_{(y\leq0)}=\psi^{BLG}e^{ik_xx}\propto e^{\lambda^{BLG}_\pm y}e^{ik_xx},\label{BLG_lambda}\\
	&&\lambda^{BLG}_\pm=\frac{1}{v}\sqrt{v^2k_x^2-\varepsilon^2-\frac{\Delta^2}{4}\pm\sqrt{\Delta^2\varepsilon^2-\gamma_1^2\left(\frac{\Delta^2}{4}-\varepsilon^2\right)}}.\nonumber
	\end{eqnarray}
	Substituting these in Eq.~(\ref{singleInt_H}) for $y\leq0$, we find that,
	\begin{eqnarray}
	&&\psi^{BLG}(k_x,\varepsilon)=
	B_+\left(\begin{array}{c}
	\alpha_+ \\ \beta_+ \\ \chi_+ \\ 1
	\end{array}\right)e^{\lambda^{BLG}_+y}+
	B_-\left(\begin{array}{c}
	\alpha_- \\ \beta_- \\ \chi_- \\ 1
	\end{array}\right)e^{\lambda^{BLG}_-y},\nonumber\\
	&&\alpha_\pm=\frac{-2\gamma_1(\Delta+2\varepsilon)(\Delta-2\varepsilon)}{X_\pm}\label{BLG_state}\\[0.5\baselineskip]
	&&\beta_\pm=\frac{4\gamma_1v(k_x+\lambda^{BLG}_\pm)(\Delta+2\varepsilon)}{X_\pm}\nonumber\\[0.5\baselineskip]
	&&\chi_\pm=\frac{(\Delta+2\varepsilon)\left[\left(\Delta-2\varepsilon\right)^2-4v^2(k_x^2-(\lambda^{BLG}_\pm)^2)\right]}{X_\pm},\nonumber\\[0.5\baselineskip]
	&&X_\pm=2v(k_x-\lambda^{BLG}_\pm)\left[\left(\Delta-2\varepsilon\right)^2-4v^2(k_x^2-(\lambda^{BLG}_\pm)^2)\right].\nonumber
	\end{eqnarray}
	Requiring continuity of the eigenstates at the BLG-2ML interface we find,
	\begin{eqnarray}
	&&\left(\begin{array}{c c c c}
	-1 & 0 \\
	-a_{-\xi_+,+} & 0 \\
	0 & -1 \\
	0 & -a_{-\xi_-,-}
	\end{array}\right)\left(\begin{array}{c}
	A_{-\xi_+,+} \\ A_{-\xi_-,-}
	\end{array}\right)\nonumber\\
	&&=\left(\begin{array}{c c c c}
	1 & 0 & -\alpha_+ & -\alpha_- \\
	a_{+\xi_+,+} & 0 & -\beta_+ & -\beta_- \\
	0 & 1 & -\chi_+ & -\chi_- \\
	0 & a_{+\xi_-,-} & -1 & -1
	\end{array}\right)\left(\begin{array}{c}
	A_{+\xi_+,+} \\ A_{+\xi_-,-} \\ B_+ \\ B_-
	\end{array}\right),\label{singleInt_matEqn_transfer}
	\end{eqnarray}
	which is nothing but the conservation condition for the current projected onto the direction perpendicular to the boundary.

\begin{figure}[!htb]
		\includegraphics[width=0.48\textwidth]{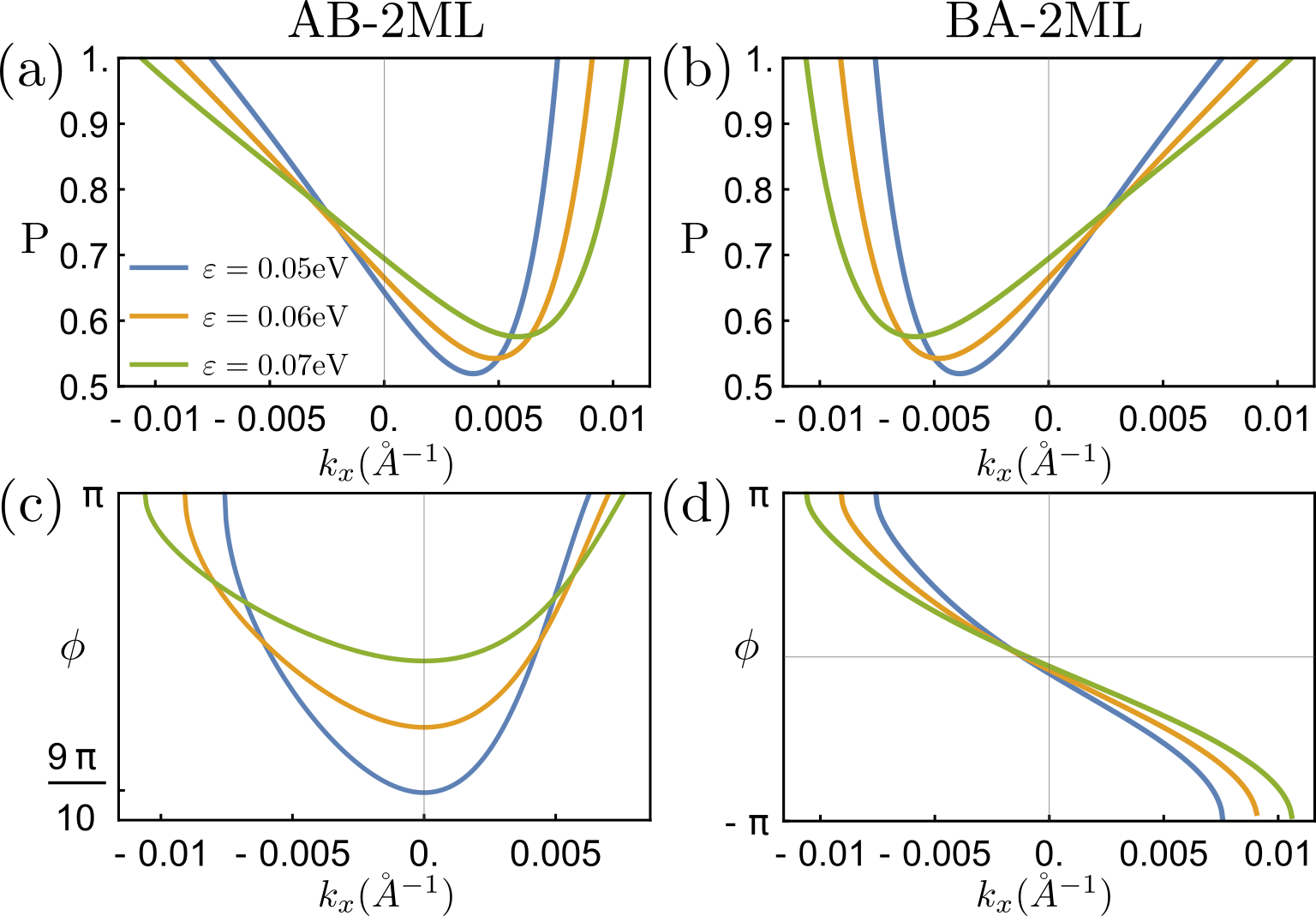}
		\caption{(a) Probability of reflection back on the same (upper) layer as the incoming wave for AB stacking within the bilayers. (b) Results for the opposite BLG stacking (BA) which exhibits identical reflection probability dependence under transformation $k_x\rightarrow-k_x$. (c-d) Phase, $\phi$, acquired upon reflection back on the same layer corresponding to panels (a-b).}
		\label{fig02:reflectionAndPhase}
	\end{figure}

	Fixing the incoming wave to reside solely on the upper monolayer ($A_{-\xi_+,+}=1$ and $A_{-\xi_-,-}=0$), we investigate how the electrons can relocate from it to the lower layer, with the amplitude of the interlayer transfer given by,
	\begin{widetext}
	\begin{equation}
	A_{+\xi_-,-}=\frac{(a_{\xi_+,+}-a_{-\xi_+,+})(\chi_+-\chi_-)}{(a_{+\xi_+,+}\alpha_+-\beta_+)(1-a_{+\xi_-,-}\chi_-)-(a_{+\xi_+,+}\alpha_--\beta_-)(1-a_{+\xi_-,-}\chi_+)}\xrightarrow[k_y\ll k_x,\delta]{}\frac{k_y}{k_x}.\label{transMatElem}
	\end{equation}
	\end{widetext}
	In the limit where momentum parallel to the interface and the monolayer on-site energy asymmetry are large, $k_y\ll k_x,\delta$, this indicates that probability of electron changing layer is small. In Fig.~\ref{fig02:reflectionAndPhase}(a) we show how the probability of reflection back on the same layer,
	\begin{equation}
	P=\frac{|A_{+\xi_+,+}|^2}{|A_{+\xi_+,+}|^2+|A_{+\xi_-,-}|^2},\label{reflectionProb}
	\end{equation}
	varies with wavevector, $k_x$, as we move around the Dirac cone at three separate energy cuts for $\delta=U=0$. In agreement with Eq.~(\ref{transMatElem}), we find that for $|k_y|\ll|k_x|$ the probability of reflection back to the same layer approaches $P=1$ (that is $A_{+\xi_-,-}\ll1$ also associated with a $\pi$ phase shift of the reflected wave), whereas peak transmission, $P\approx0.5$, onto the second layer occurs for waves incident at angle $\pi/3$.
	
	\subsection{Evanescent interface states}\label{singleIntStateSec}
	
\begin{figure}[!htb]
		\includegraphics[width=0.4\textwidth]{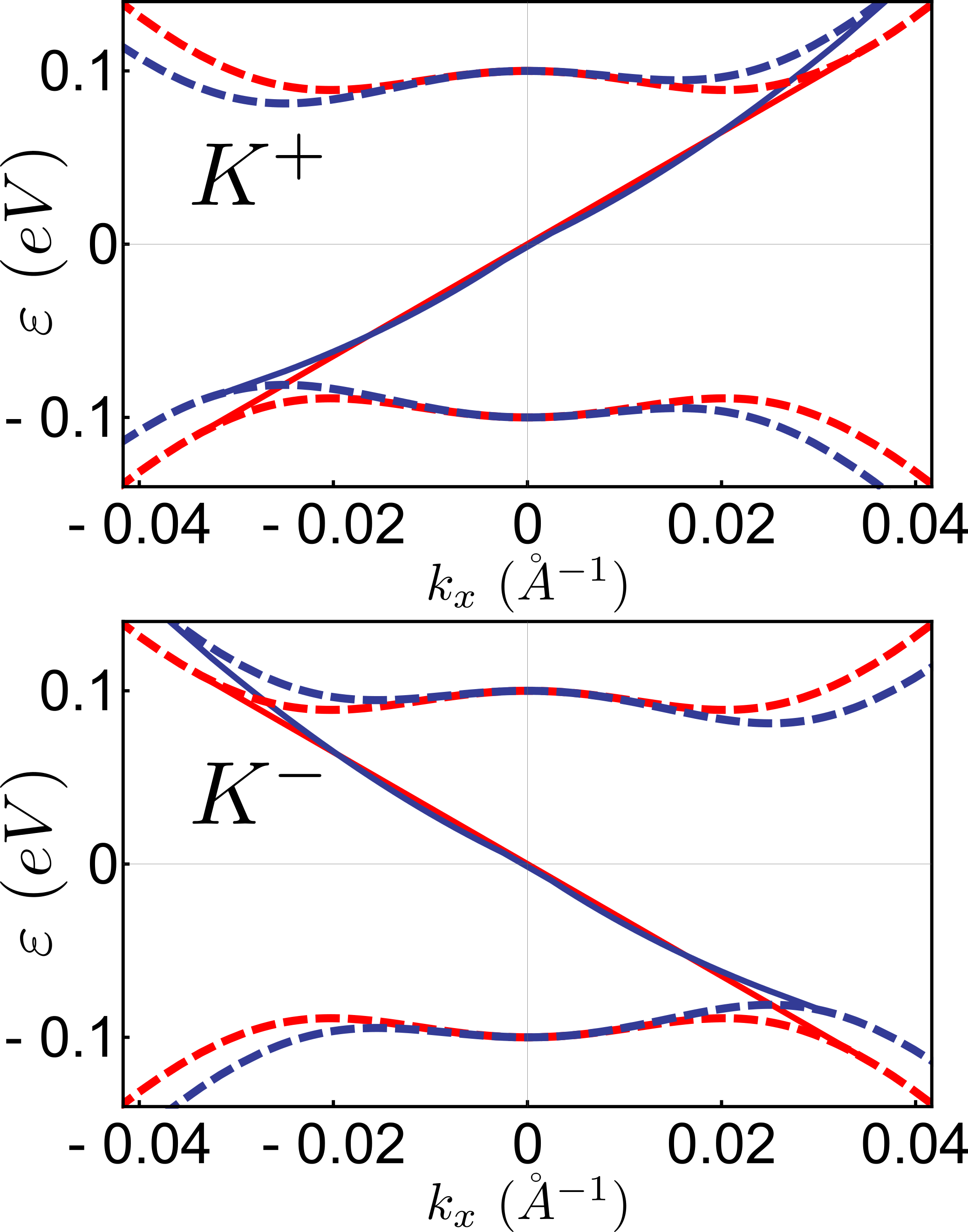}
		\caption{Interface states (solid red curves) and low energy BLG bands (dashed red curves) at $y=0$ for $\Delta=0.2$~eV and $\delta=U=0$ in the $K^+$ (top) and $K^-$ (bottom) valley. Blue curves depict the bands for non-zero skew inter-layer coupling, $\gamma_3$, for a delamination along the zig-zag direction, illustrating the negligible effect that this additional hopping term has on the spectrum at a delamination edge.}
		\label{fig03:singleIntModes}
	\end{figure}

	In addition to scattered waves, a BLG-2ML interface supports evanescent modes over the entire energy range $|\varepsilon|<\varepsilon_*$. These are described by Eq.~(\ref{BLG_state}) in the BLG region, and by Eq.~(\ref{2ML_state}) in the 2ML delamination, but, now, with $q_{s,l}=-i\lambda^{2ML}_{s,l}$ where,
	\begin{equation}
	\lambda^{2ML}_{s,l}=s\frac{1}{v}\sqrt{v^2k_x^2-\left(\varepsilon-U-l\frac{\delta}{2}\right)^2},
	\end{equation}
	are real-valued decay rates. Continuity of these evanescent wavefunctions across the interface (selecting e.g, $s=-1$) produces matching condition,
	\begin{equation}
	\underbrace{\left(\begin{array}{c c c c}
	\alpha_+ & \alpha_- & -1 & 0 \\
	\beta_+ & \beta_- & -a_{-,+} & 0 \\
	\chi_+ & \chi_- & 0 & -1 \\
	1 & 1 & 0 & -a_{-,-}
	\end{array}\right)}_\mathlarger{\mathcal{D}(\varepsilon,k_x)}\left(\begin{array}{c}
	B_+ \\ B_- \\ A_{-,+} \\ A_{-,-}
	\end{array}\right)=0.\label{singleInt_matEqn_boundary}
	\end{equation}
	To find its solution, we have to require that,
	\begin{equation}
	\det\mathcal{D}(\varepsilon,k_x)=0\nonumber
	\end{equation}
	which sets the dispersion relation $\varepsilon(k_x)$ for the evanescent modes. Figure \ref{fig03:singleIntModes} shows the results of solving numerically for these 1D states localised near the BLG-2ML interface for $\Delta=0.2$~eV and $\delta=0$. Having noticed an almost linear dispersion of such states, we also find that the dispersion of evanescent modes is almost linear for arbitrary values of all electrostatically controlled parameters in our theory, assuming that $vk_x,\Delta,\delta,U\ll\gamma_1$,
	\begin{equation}
	\varepsilon\approx\eta\sqrt{\frac{\Delta}{\gamma_1}}v k_x,\label{singleInt_boundaryAnalytic}
	\end{equation}
	and modes in the opposite valleys, $\eta=\pm$, propagate in opposite directions. Exchanging the interlayer stacking configuration in the bilayer region (AB$\rightarrow$BA) or swapping the ordering of the two regions (2ML on the left and BLG on the right) results in mirror-reflected dispersions equivalent to those illustrated in Fig.~\ref{fig03:singleIntModes}.
	
	\section{Electronic Spectrum of a Delamination in BLG}\label{doubleIntSec}

	Here, we analyse the electronic spectrum of a delamination (2LG) between two bilayer regions to the left ($y\leq0$) and right ($y\geq W$) hand sides (see Fig.~\ref{fig01:delaminationSketch}) for the same (AB-2ML-AB) and opposite (AB-2ML-BA) stacking on the two sides.
	
	\subsection{AB-2ML-BA Stacking}
	
	Depending on the fabrication process generating the delamination, the layers in a BLG may be shifted from one Bernal stacking configuration to another, producing the inverted (BA vs AB) stacking on the opposite side of the delamination. The resulting deformation, of the order of one carbon-carbon bond length of graphene ($a_{cc}=1.42$~\AA), is then absorbed by a weak strain/shear of the delaminated monolayers. The effect of weak strain homogeneous over a narrow stripe of the upper/lower monolayer in the 2ML part of the structure consists of an addition of a vector potential term to the Dirac equation in the monolayers \cite{rainis_PRB_2011}, producing small shifts in the wave numbers and energies of modes guided by the delamination. In the following, we neglect these small ($\sim a/W$) shifts and focus on the qualitative change in the delamination spectrum brought about by the stacking order: its gapless character.
	
\begin{figure*}
		\includegraphics[width=\textwidth]{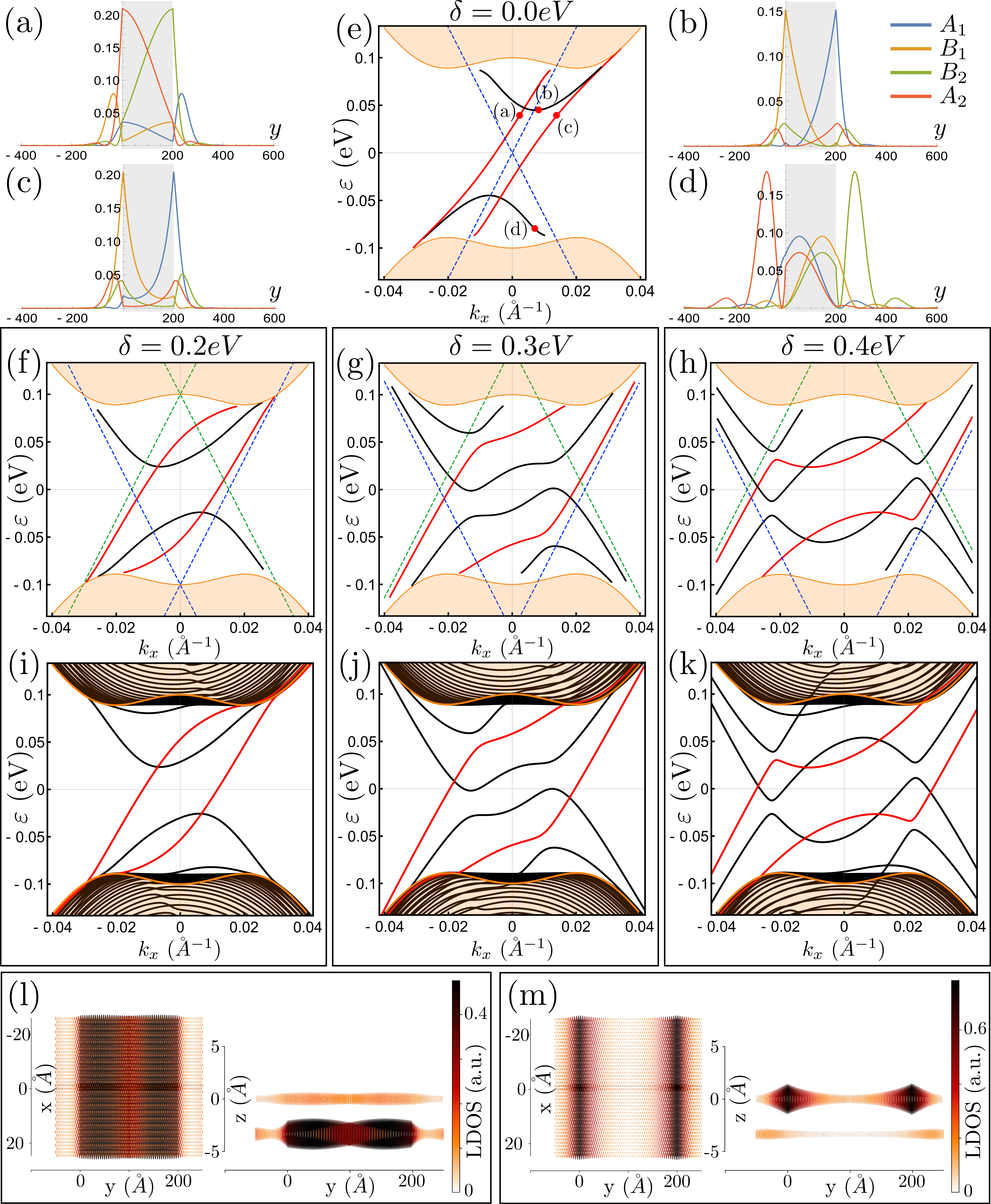}
		\caption{(a-d) Wavefunction across the 2ML channel for specific energy and momenta (indicated on plot (e)) in the sub-lattice basis. (e-h) Permitted energy bands calculated from the continuum model within the bilayer gap, whilst (i-k) demonstrate the results of an equivalent TB calculation approach including continuum bands which extend above the bilayer band edge. Interface modes (red curves) are identified crossing between the low-energy bilayer band edges (orange shaded regions). Plotted in green (blue) dashed lines are the conical dispersions of the upper (lower) monolayer with energy difference, $\delta$, between their Dirac points. (l,m) Top and side profiles of the LDOS for the same parameters as in (c \& d) Calculations performed for $W=200$~\AA, $\Delta=0.2$~eV and $U=0$ around the $K^+$ valley.}
		\label{fig04:delaminationSolns_ABBA}
	\end{figure*}

	The continuous model for the system with AB-2ML-BA stacking, written in basis $(A_1,B_1,B_2,A_2)^T$, is described by Hamiltonian,
	\begin{equation}
	\hat{{\cal H}}_\eta=\left(\begin{array}{c c c c}
	V_+(y) & v\hat{\pi}^\dagger & \Theta(-y)\gamma_1 & 0\\
	v\hat{\pi} & V_+(y) & 0 & \Theta(y-W)\gamma_1\\
	\Theta(-y)\gamma_1 & 0 & V_-(y) & v\hat{\pi}\\
	0 & \Theta(y-W) \gamma_1 & v\hat{\pi}^\dagger & V_-(y)
	\end{array}\right),\label{doubleInt_H_BA}
	\end{equation}
	where on-site energies on the upper (+) and lower (-) layers are,
	\begin{equation}
	V_\pm(y)=\begin{cases}
	\pm\frac{\Delta}{2} & \text{if } y\leq0 \text{ or } y\geq W \\
	U\pm\frac{\delta}{2} & \text{if } 0<y<W
	\end{cases}.\label{doubleInt_cases_V}
	\end{equation}
	For $y<0$ and $0<y<W$, the wavefunctions retain the form given in Eqs.~(\ref{2ML_state} \& \ref{BLG_state}), whereas for $y>W$ (BA stacked BLG) parameters in $\psi^{BLG}$ in Eq.~(\ref{BLG_state}) should be substituted with,
	\begin{eqnarray}
	&&\tilde{\alpha}_\pm=\frac{4\gamma_1v(\Delta+2\varepsilon)(k_x-\lambda^{BLG}_\pm)}{X_\pm}\label{BA_BLG_elems}\\[0.5\baselineskip]
	&&\tilde{\beta}_\pm=-\frac{2\gamma_1(\Delta+2\varepsilon)(\Delta-2\varepsilon)}{X_\pm}\nonumber\\[0.5\baselineskip]
	&&\tilde{\chi}_\pm=\frac{2v(k_x+\lambda^{BLG}_\pm)\left[\left(\Delta-2\varepsilon\right)^2-4v^2(k_x^2-(\lambda^{BLG}_\pm)^2)\right]}{X_\pm},\nonumber\\[0.5\baselineskip]
	&&\tilde{X}_\pm=(\Delta+2\varepsilon)\left[\left(\Delta-2\varepsilon\right)^2-4v^2(k_x^2-(\lambda^{BLG}_\pm)^2)\right]\nonumber.
	\end{eqnarray}
	Wavematching conditions applied at both interfaces, $\psi(y\rightarrow0^-)=\psi(y\rightarrow0^+)$ and $\psi(y\rightarrow W^-)=\psi(y\rightarrow W^+)$, result in,
	\begin{widetext}
		\begin{equation}
		\underbrace{\left(\begin{array}{c c c c c c c c}
			\alpha_+ & \alpha_- & 1 & 1 & 0 & 0 & 0 & 0 \\
			\beta_+ & \beta_- & a_{+,+} & a_{-,+} & 0 & 0 & 0 & 0 \\
			\chi_+ & \chi_- & 0 & 0 & 1 & 1 & 0 & 0 \\
			1 & 1 & 0 & 0 & a_{+,-} & a_{-,-} & 0 & 0 \\
			0 & 0 & e^{\lambda^{MLG}_{+,+}W} & e^{\lambda^{MLG}_{-,+}W} & 0 & 0 & \tilde{\alpha}_+e^{-\lambda^{BLG}_+W} & \tilde{\alpha}_-e^{-\lambda^{BLG}_-W} \\
			0 & 0 & a_{+,+}e^{\lambda^{MLG}_{+,+}W} & a_{-,+}e^{\lambda^{MLG}_{-,+}W} & 0 & 0 & \tilde{\beta}_+e^{-\lambda^{BLG}_+W} & \tilde{\beta}_-e^{-\lambda^{BLG}_-W} \\
			0 & 0 & 0 & 0 & e^{\lambda^{MLG}_{+,-}W} & e^{\lambda^{MLG}_{-,-}W} & \tilde{\chi}_+e^{-\lambda^{BLG}_+W} & \tilde{\chi}_-e^{-\lambda^{BLG}_-W} \\
			0 & 0 & 0 & 0 & a_{+,-}e^{\lambda^{MLG}_{+,-}W} & a_{-,-}e^{\lambda^{MLG}_{-,-}W} & e^{-\lambda^{BLG}_+W} & e^{-\lambda^{BLG}_-W}
			\end{array}\right)}_\mathlarger{{\tilde{\mathcal{D}}(\varepsilon,k_x)}}
		\left(\begin{array}{c}
		B_+ \\ B_- \\ A_{+,+} \\ A_{-,+} \\ A_{+,-} \\ A_{-,-} \\ \tilde{B}_+ \\ \tilde{B}_- \end{array}\right)=0.\label{doubleInt_matEqn}
		\end{equation}
	\end{widetext}
	As with the single interface system, we require that,
	\begin{equation}
	\det\tilde{\mathcal{D}}(\varepsilon,k_x)=0.\nonumber
	\end{equation}
	Solving this equation numerically, we generate dispersion curves $\varepsilon(k_x)$ shown in Fig.~\ref{fig04:delaminationSolns_ABBA} which represent the main features of the delamination spectra.
	
	Figures \ref{fig04:delaminationSolns_ABBA} illustrate the bands (e-h) and wavefunctions (a-d) for a range of monolayer band offsets, calculated from the continuum model. These plots show the spectra of electrons in valley $K^+$. Dispersions in the valley $K^-$ can be obtained using the time reversal, $\varepsilon(K^-,k_x)=\varepsilon(K^+,-k_x)$. Wavefunctions for the interface states are displayed in Figs.~\ref{fig04:delaminationSolns_ABBA}(a,c), which demonstrate localisation of these valley-polarised modes at both interfaces. Note, that although an increasing number of bands become available as we open up the 2ML gap, $\delta$, there are only these two interface states which span the entire region. Therefore, tuning the Fermi level of the system such that it lies close to zero will select states corresponding to one-dimensional ballistic channels propagating simultaneously along each interface. The gap, $\varepsilon_g$, between the lowest `conduction' and `valence' bands is shown to decrease as $\varepsilon_g\propto W^{-1}$ with delamination width (see Appendix \ref{gapsWithW}).
	
	We also employ a numerical tight binding (TB) approach, by modelling the considered structure as a semi-infinite zig-zag BLG nanoribbon of lattice sites with a delaminated region separating two BLG regions with opposite stacking. We incorporate a mismatch of one carbon-carbon bond-length into the widths of the layers in the 2ML region \cite{koshino_PRB_2013}. Modifying the standard TB Hamiltonian for BLG to include the desired structure produces,
	\begin{equation} 
	\begin{split}
	\hat{\mathcal{H}} =& -\sum_l\sum_{\braket{i,j}}(\gamma_0 c_{l,i}^{\dagger}c_{l,j}) \\
	&-\sum_{\braket{i}}\left\{\left(\Theta(-y_i)+\Theta(y_i-W)\right)\gamma_1 c_{1,i}^{\dagger}c_{2,i} + \text{h.c}\right\} \\
	&+\sum_i V_{\pm}(y_i) c_i^{\dagger}c_i,
	\end{split}	\label{Eqn:TB_Hamiltonian}
	\end{equation}
	with $\gamma_0=3.1$ the intralayer nearest neighbour coupling, $\gamma_1=0.39$~eV the interlayer nearest neighbour coupling, and $c_{l,i}^{\dagger}\ (c_{l,i})$ being the creation (annihilation) operator for electrons at site $\vec{R}_i=(x_i,y_i)$ in layer $l$. On-site potentials, $V_{+}(y_i) (V_{-}(y_i))$, on the upper (lower) layer are given by Eq.~(\ref{doubleInt_cases_V}). The sum in the first line runs over all nearest neighbours, $\braket{i,j}$, the second line runs over all the coupled dimer sites $\braket{i}$, and the final term sums over all lattice sites, $i$. The difference between the AB and BA regions is taken into account as a change of the 'dimer' sites, meaning that the coupling is present between sites $A_1$-$B_2$ ($\Theta(-y_i)$), and $B_1$-$A_2$ ($\Theta(y_i-W)$) respectively.
	
	For numerical diagonalisation, we used 2000\AA\ for the total width of the AB-2ML-BA nanoribbon along the $y$ axis, and zig-zag edges. To prevent states localised along these terminating edges \cite{mccann_JPhysCondMat_2004,akhmerov_PRB_2008,slizovskiy_PRB_2017} from obscuring the states in the delaminated region, we apply a large positive (negative) on-site potential on the edge atoms which pushes these states to higher (lower) energies.
	
	Using TB Hamiltonian (\ref{Eqn:TB_Hamiltonian}) we find the spectrum of states guided by the delamination by solving the equation, $\left(\hat{\mathcal{H}}(k_x)-\varepsilon(k_x)\hat{\mathcal{I}}\right)\Psi=0$, for different values of wavevector, $k_x$. Additionally, using the Kernel Polynomial Method \cite{weise_RevModPhys_2006} implemented in the Pybinding package \cite{moldovan_pybinding_2017} we investigate the density of states (DOS) and local DOS (LDOS).
	
	The spectra found using the TB approach are shown in Fig.~\ref{fig04:delaminationSolns_ABBA}(i-k) for the same parameters as in continuum theory [panels (f-h)]\cite{footNote1}. As with the AB-2ML-AB system, the delamination exhibits both localised channels along the interface and standing wave modes across the delaminated monolayers, but the interface states now span the full range of the BLG gap (identified in red in (e-k)). Note that the spectra obtained by the two methods coincide in all details, including all avoided and non-avoided crossings between interface states and standing wave state. Also confirmed by both calculation methods is that the change in stacking order breaks the $\varepsilon(k_x)=\varepsilon(-k_x)$ symmetry for the states in one valley and that $\varepsilon_{K^+}(k_x)=\varepsilon_{K^-}(-k_x)$.

	\subsection{AB-2ML-AB Stacking}
	
	Having established a good agreement between the properties of ballistic electron channels evaluated using the TB approach and wavematching, we study a delamination with the same stacking in the outer BLG regions using only the continuous theory, with Hamiltonian,
	\begin{equation}
	\hat{{\cal H}}_\eta=\left(\begin{array}{c c c c}
	V_+(y) & v\hat{\pi}^\dagger & \Phi\gamma_1 & 0\\
	v\hat{\pi} & V_+(y) & 0 & 0\\
	\Phi\gamma_1 & 0 & V_-(y) & v\hat{\pi}\\
	0 & 0 & v\hat{\pi}^\dagger & V_-(y)
	\end{array}\right).\label{doubleInt_H_AB}
	\end{equation}
	Here $\Phi=\Theta(-y)+\Theta(y-W)$ and $V_\pm(y)$ take the same values as in Eq.~(\ref{doubleInt_cases_V}).	Electron states for $0<y<W$ are described by Eq.~(\ref{2ML_state}), where wavevectors $q_{s,l}$ may take both real and imaginary values. For energies $|\varepsilon|<\varepsilon_*$, states in the left bilayer remain evanescent, Eq.~(\ref{BLG_state}), with states in the right hand side bilayer obtained by substitution $y\rightarrow W-y$. 
	
\begin{figure*}
		\centering
		\includegraphics[width=\textwidth]{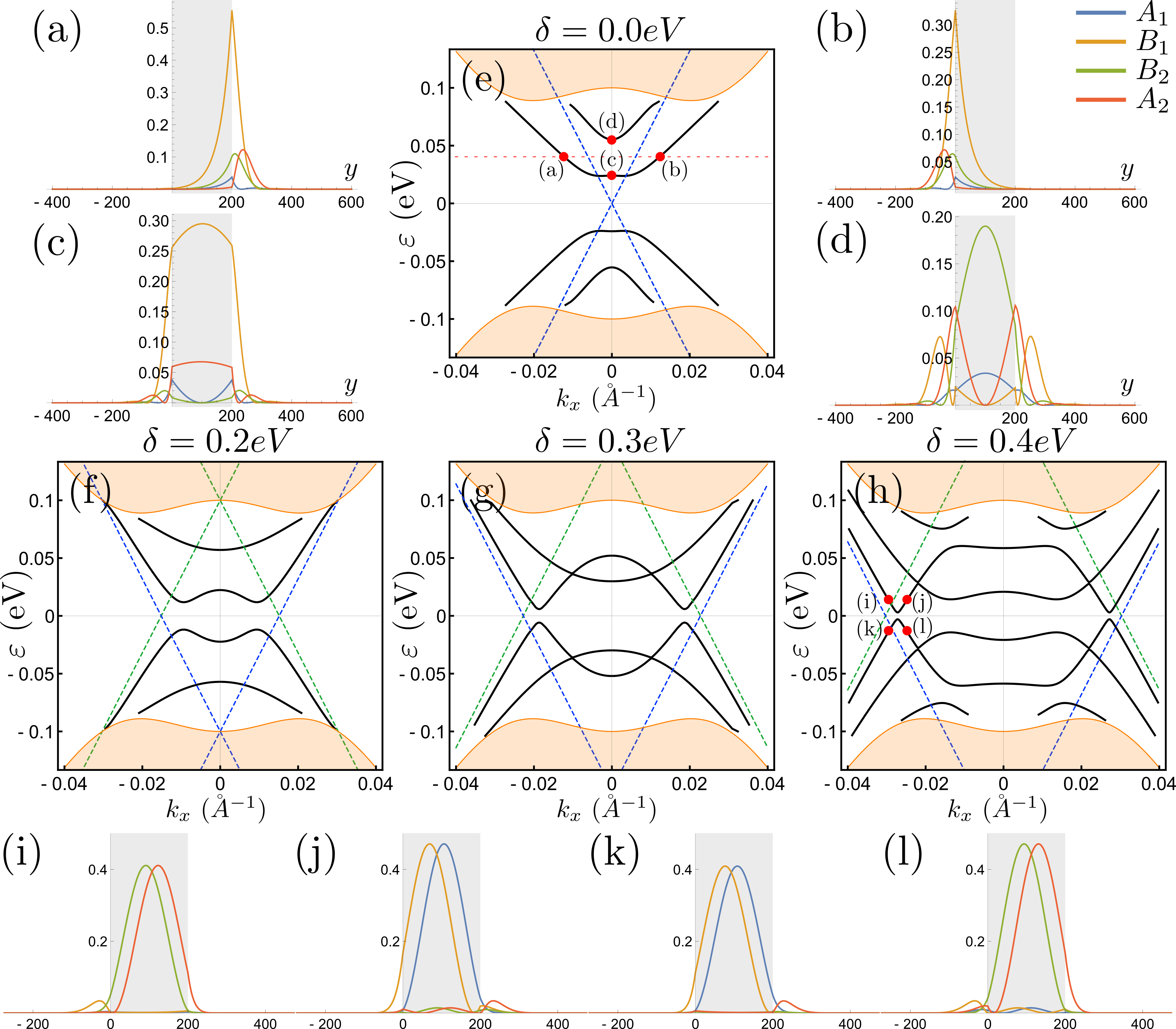}
		\caption{(a-d,i-l) Electron wavefunctions across the delamination (shaded grey region) for specific energy and momenta indicated in panels (e) and (h) in sub-lattice basis $(A_1,B_1,B_2,A_2)^T$. (e-h) Energy structure of the delamination for increasing interlayer energy gap ($\delta$) within the monolayer region. Green (blue) dashed lines denote the upper (lower) monolayer Dirac cones in the 2ML region, whilst the orange region denotes the low-energy bulk bilayer band edge. Calculations are for $W=200$~\AA, $\Delta=0.2$~eV and $U=0$ around the $K^+$ valley.}
		\label{fig06:delaminationSolns_ABAB}
	\end{figure*}

	Matching these wavefunctions at the two interfaces we recover Eq.~(\ref{doubleInt_matEqn}), where $\alpha_\pm$, $\beta_\pm$ and $\chi_\pm$ are defined in Eq.~(\ref{BLG_state}), and $\tilde{\alpha}_\pm=\alpha_\pm(-\lambda^{BLG}_\pm)$, $\tilde{\beta}_\pm=\beta_\pm(-\lambda^{BLG}_\pm)$ and $\tilde{\chi}_\pm=\chi_\pm(-\lambda^{BLG}_\pm)$ such that the same AB-stacked BLG states decay away to the right of the delamination. As with the AB-2ML-BA system we require that $\det\tilde{\mathcal{D}}=0$, producing dispersions across the delamination shown in Fig.~\ref{fig06:delaminationSolns_ABAB}.
	
	These spectra are formed by the hybridisation of evanescent modes coming from the opposite edges of the delamination and standing waves bouncing between the edges. For wider delaminations or larger $\delta$, we find more `sub-bands' that can fit the BLG asymmetry gap. Higher sub-bands correspond to bouncing modes within the monolayers. Note that the spectra shown in Fig.~\ref{fig06:delaminationSolns_ABAB} are gapped. As with the AB-2ML-BA system, the values of energy gaps between bands exhibits $\varepsilon_g\propto(W+W_0)^{-1}$ dependence with delamination width (see Appendix \ref{gapsWithW})
	
	The gaps in the delamination spectrum also depend on the interlayer asymmetry, $\delta$, which is controlled by the displacement field $E_z$. Wavefunctions in Panels (i-l) demonstrate that, for large $\delta$, electronic states just above and below the avoided crossings are localised on different delaminated monolayers. The crossing occurs at momentum $k_x\approx\delta/2v$, and the standing wave states within the delamination have $k_y=\pi/W\ll\delta/2v$. Using Eq.(\ref{transMatElem}) for the interlayer transmission amplitude we estimate for the size of the gap,
	\begin{equation}
	\varepsilon_g\propto|A_{+\xi_-,-}|\approx\frac{2v\pi/W}{\delta},
	\end{equation}
	which is in agreement with our numerical data.
	
\begin{figure}[!ht]
		\includegraphics[width=0.48\textwidth]{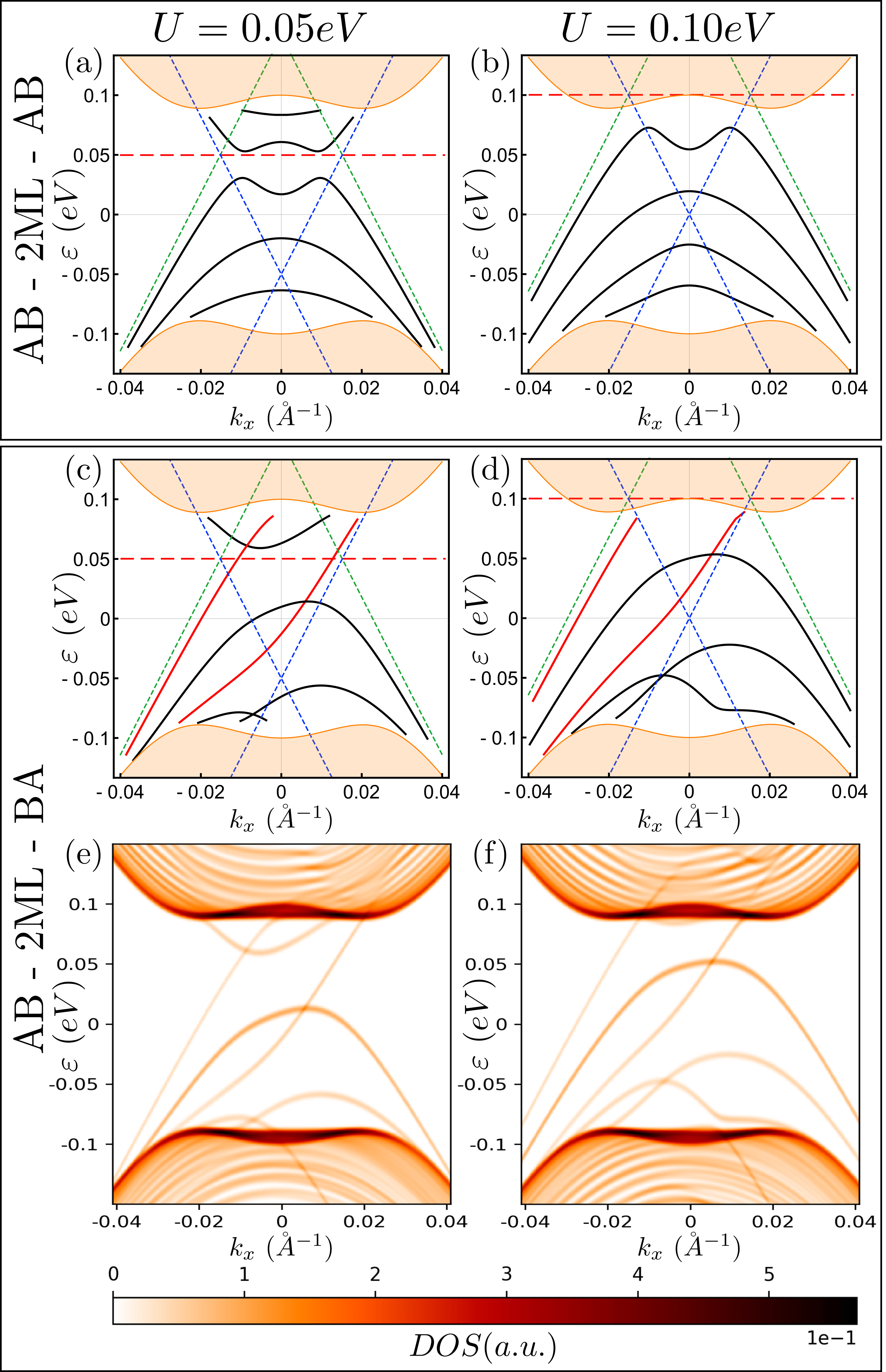}
		\caption{Inter-band states for different symmetry-breaking energy offsets, $U$, for both AB-2ML-AB (a,b) and AB-2ML-BA (c-f) stacking with $W=200$~\AA\ and $\Delta=\delta=0.2$~eV. In figures (a-d), calculated using the continuum model, the red dashed line indicates the 2ML region energy mid-point and green/blue dashed lines indicate the positions of the bulk 2ML Dirac cones. (e,f) Density of states of the system calculated from the TB model.}
		\label{fig05:energyAsym}
	\end{figure}

	Additionally, we studied how an off-set, $U$, between the BLG and 2ML regions affects the energy dispersions of both the AB-2ML-AB and AB-2ML-BA structures. Figure \ref{fig05:energyAsym} depicts this for two different offsets in both the AB-2ML-AB and AB-2ML-BA interlayer stacking: in (a-d) are bands calculated using the continuum model, and in (e-f) are bands found using the TB model in the form of the density of state (DOS) maps. It shows that an increase in either $U$ or $\delta$ (or both) brings more sub-bands into the delamination spectrum inside the BLG gap.
	
	\section{Delamination Spectrum in a Perpendicular Magnetic Field}\label{magFieldSec}
	
	To investigate the effects of a perpendicular magnetic field on the interface states, we use the TB model and applying a Peierls substitution to the in-plane coupling terms in Eq.~(\ref{Eqn:TB_Hamiltonian}),
	\begin{equation}
		\begin{split}
			\hat{\mathcal{H}} =& -\sum_l\sum_{\braket{i,j}}(\gamma_0e^{i2\pi\Phi_{ij}/\Phi_0} {c_{l,i}}^{\dagger}c_{l,j})\\ 
			&-\sum_{\braket{i}}\left\{\left(\Theta(y_i)+\Xi(y_i)\right)\gamma_1 c_{1,i}^{\dagger}c_{2,j} + \text{h.c}\right\}\\
			&+\sum_i V_{\pm}(y_i) c_i^{\dagger}c_i. 
		\end{split}
		\label{Eqn:TB_Hamiltonian_magn_field}
	\end{equation}
	Here $\Phi_0=\frac{h}{e}$ is the magnetic flux quantum and $\Phi_{ij}=\int\vec{A}\vec{dl}$ is the flux accumulated between atomic sites $i$ and $j$ due to the external magnetic field. Using the Landau gauge, $\vec{A}=By\hat{x}$, we find, $\Phi_{ij}=0.5B(y_j+y_i)(x_j-x_i)$.
	
	\begin{figure}
		\includegraphics[width=0.48\textwidth]{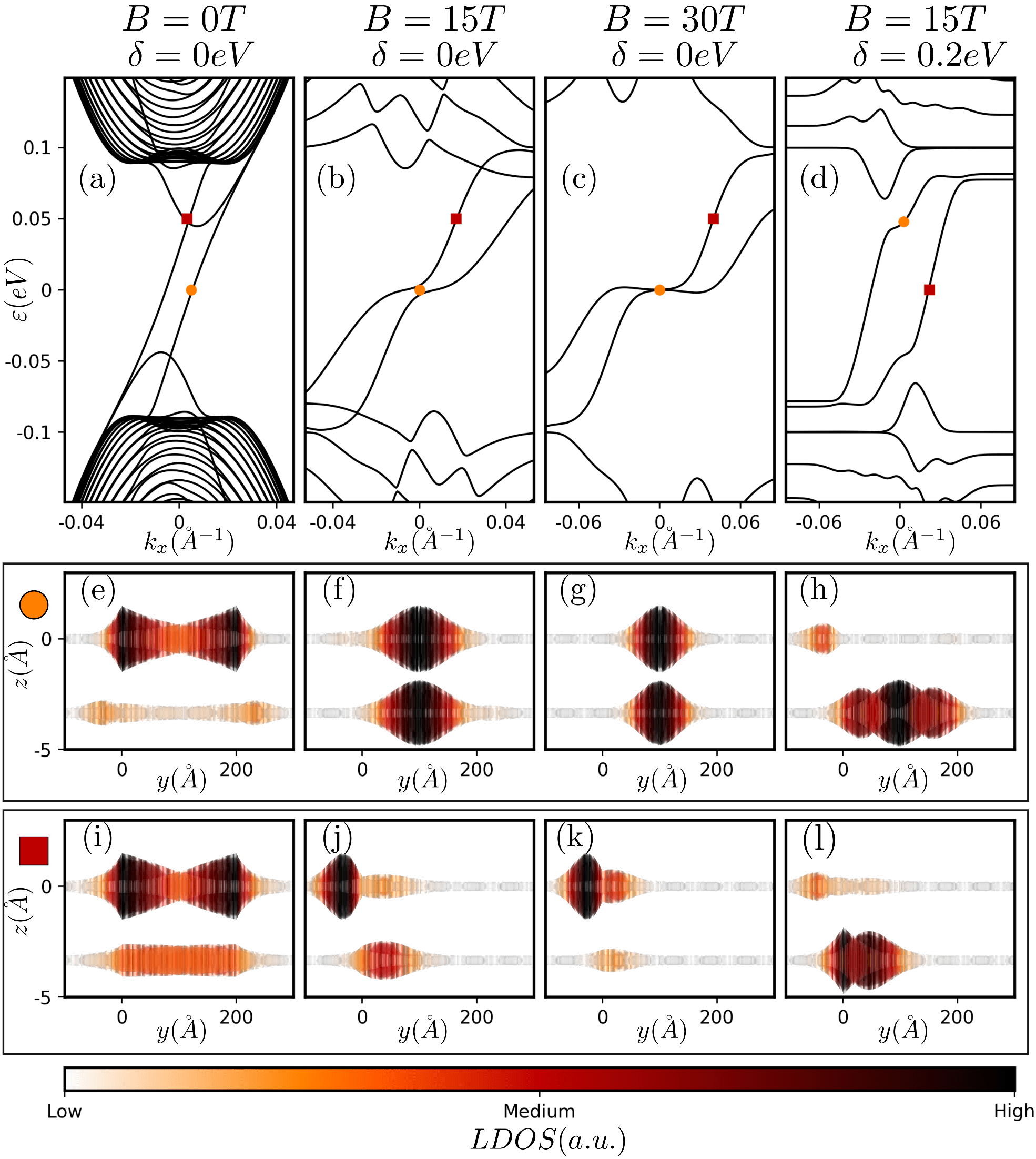}
		\caption{(a-d) Band structure and (e-l) LDOS maps at specific values of energy and wavevector for the nanoribbon in the presence of a perpendicular magnetic field. Insets (e-h) and (i-l) depict the distribution of the states at the value of $\varepsilon$ and $k_x$ marked on the band structure plots with yellow circles and red squares respectively. Calculations are for $W=200$~\AA, $\Delta=0.2$~eV and $U=0$ in the vicinity of the $K^+$ valley.}
		\label{fig07:magField}
	\end{figure}
	
	Figs.~\ref{fig07:magField}(a-d) show that a perpendicular magnetic field increases\cite{yin_natComms_2016} the visibility of the interface states. The wavefunction of the standing wave states are delocalised over the relatively wide delaminated region and therefore readily form Landau levels when the radius of the cyclotron orbit of monolayer graphene drops below half the delamination width ($r_0\sim l_B = \sqrt{\hbar/eB} < W/2$). 
	For a $W=200$~\AA\ structure (Figs.~\ref{fig07:magField}(a-d)), this corresponds to a magnetic field of $B\approx7$ Tesla, and results in the energies of these states being pushed out of the energy range of the bandgap in bilayer graphene. In contrast, the interface states are less susceptible to Landau level formation as their unidirectional propagation and confinement near the edges inhibits the formation of cyclotron orbits.
	Nevertheless, as the magnetic field strength increases from $0$ to $30$~T, a clear precursor to a zeroth Landau level is seen in the flattening of the dispersions near $k_x=\epsilon=0$ and a concomitant localisation of the wavefunction in the centre of the delaminated region (compare \ref{fig07:magField} (a-c,e-g)). Similar flattening of the interface bands is found for a finite interlayer shift near $k_x=0$, $\epsilon=\pm0.045$~eV (e.g yellow dot, \ref{fig07:magField} (d)). The energies of these features are well approximated by the first Landau level energy counted from the Dirac cone in the appropriate layer, $\epsilon =\pm(v\sqrt{2}/l_B - \delta/2)=0.041$~eV, and the corresponding wavefunction (\ref{fig07:magField} (h)) displays characteristics of the first graphene Landau level.
	
	For energies away from these band flattenings (e.g. red dots in \ref{fig07:magField} (b-d)) the wavefunction is found to be pushed towards the left interface (\ref{fig07:magField} (j-l)), rather than in the centre of the delamination as is found for the Landau level like features \ref{fig07:magField} (f-h). This behaviour can be interpreted as the result of a Lorentz force, $e\vec v\times\vec B$, for an electron with velocity in the x-direction. As the velocity of the interface states is reversed for the other ($K^-$) valley the corresponding states there are pushed towards the opposite edge. This produces a spatial separation of the wavefunctions in the two valleys and consequently an enhanced robustness of these states against intervalley scattering disorder \cite{li_natNano_2016}.	

	\section{Discussions}\label{discussionsSec}
	
	In this paper, we have studied the electronic properties of delaminations in bilayer graphene with a gate induced bandgap. Starting with a single boundary between a bilayer graphene region and two decoupled monolayer graphene sheets, we have shown that there exist evanescent states localized at the interface which span the bilayer energy gap and have opposite carrier velocity in each of graphene's two valleys. In a delamination stripe with opposite stacking of the outer BLG, these evanescent states give rise to the gapless channels counterpropagating in opposite valleys. The delamination, with both AB-2ML-BA and AB-2ML-AB stacking, also support channels produced by bouncing modes in the two delaminated monolayers. The number of such modes increases with delamination width, $W$, as well as with the transverse electric field, $E_z$, which controls the difference, $\delta=eE_zd'$, between on-layer electron energies in the delamination.
	
\begin{figure}
		\includegraphics[width=0.35\textwidth]{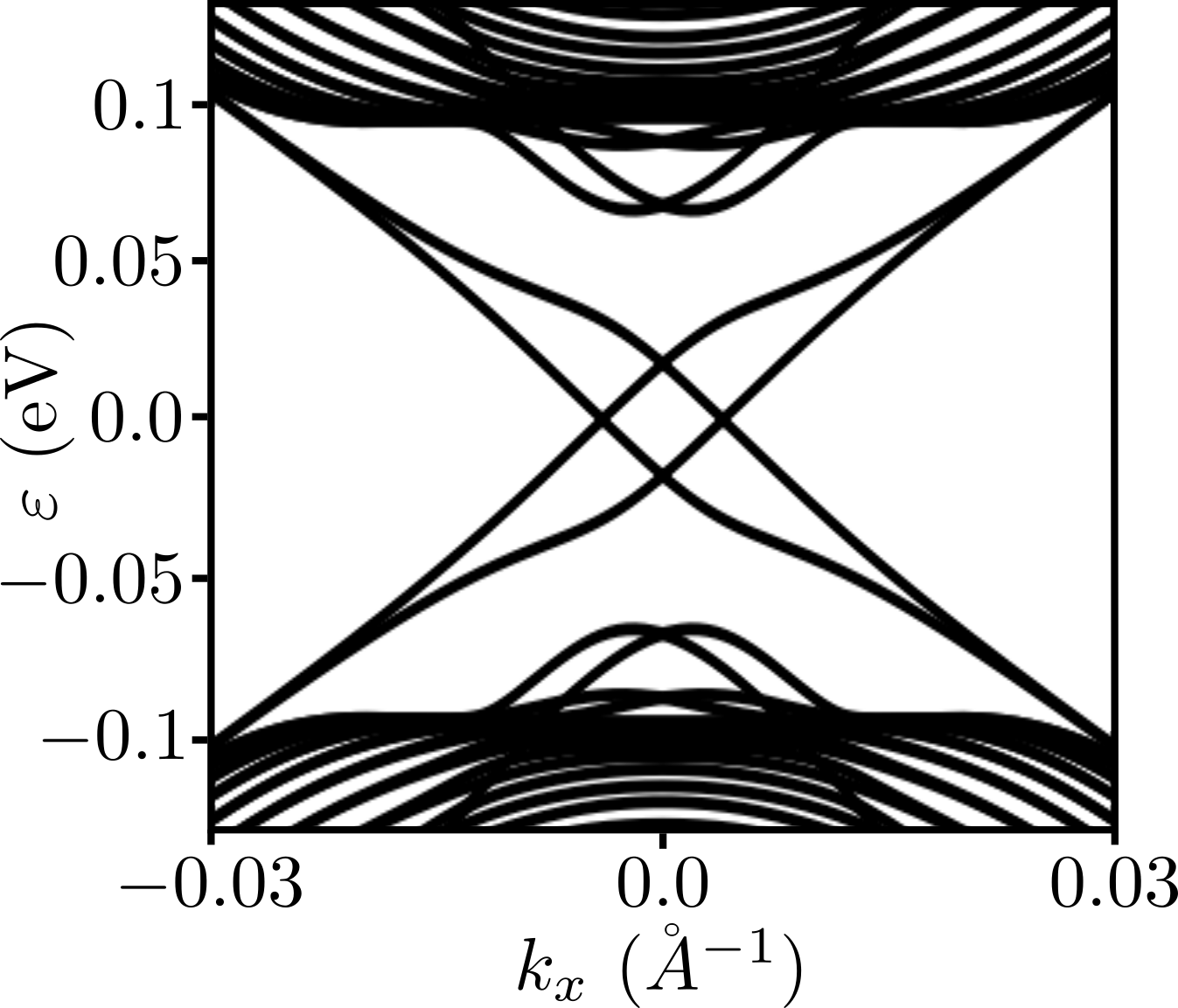}
		\caption{Energy dispersion for an AB-2ML-BA delamination with armchair edges, calculated using the TB model for $\Delta=0.2$~eV, $\delta=U=0$ and $W=200$\AA.}
		\label{fig08:armchair}
	\end{figure}

	The above results have been obtained for the delamination with an almost arbitrary crystallographic direction, except for the orientation where the delamination edge exactly coincides with the armchair direction in graphene. In the latter case, the armchair edge mixes states in graphene's two valleys and the spectrum of an AB-2ML-BA delamination, obtained by TB model calculation and shown in Fig.~\ref{fig08:armchair}, has modes inside the BLG gap which are not protected against anticrossing by the valley structure of electron states. Such states seem to retain linear dispersion similar to the case of a generic orientation of delamination edges.
	
	Another point to mention is related to the role of $\gamma_3$ hopping terms in the TB model, which were neglected in modelling the electronic spectra. Those terms produce \cite{mccann_PRL_2006} only weak trigonal warping effects in BLG and can be neglected, but also (together with the variation of $\gamma_1$ coupling along the delamination edge) may generate a source of scattering at the exact delamination edge with arbitrary (different from exact zig-zag) orientation. The analysis of such disorder will be a subject for a separate study.
	
	\begin{acknowledgements}
		This work was funded by EPSRC via EPSRC Grand Engineering Challenges grant EP/N010345, the Manchester NOWNANO CDT EP/L-1548X, the Flemish Science Foundation (FWO-Vl), European Graphene Flagship project, ERC Synergy grant Hetero2D and FLAG-ERA project TRANS2DTMD. The author would like to acknowledge useful discussions with M. Zarenia, S.Slizovskiy E. McCann, and K. Novesolov.
	\end{acknowledgements}
	
	\appendix
	
	\section{Energy Gaps in Delaminated Systems}\label{gapsWithW}
	
	The number of available states within the BLG gap and their corresponding energies are highly sensitive, not only to the electrostatic parameters (as demonstrated in Figs.~(\ref{fig04:delaminationSolns_ABBA}\&\ref{fig06:delaminationSolns_ABAB})) but also to the width of the stacked monolayer channel. Decreasing the width, $W$, of the delamination opens up larger energy gaps between subsequent bands and pushes them out of our energy range of interest.
	
	\begin{figure}
		\includegraphics[width=0.48\textwidth]{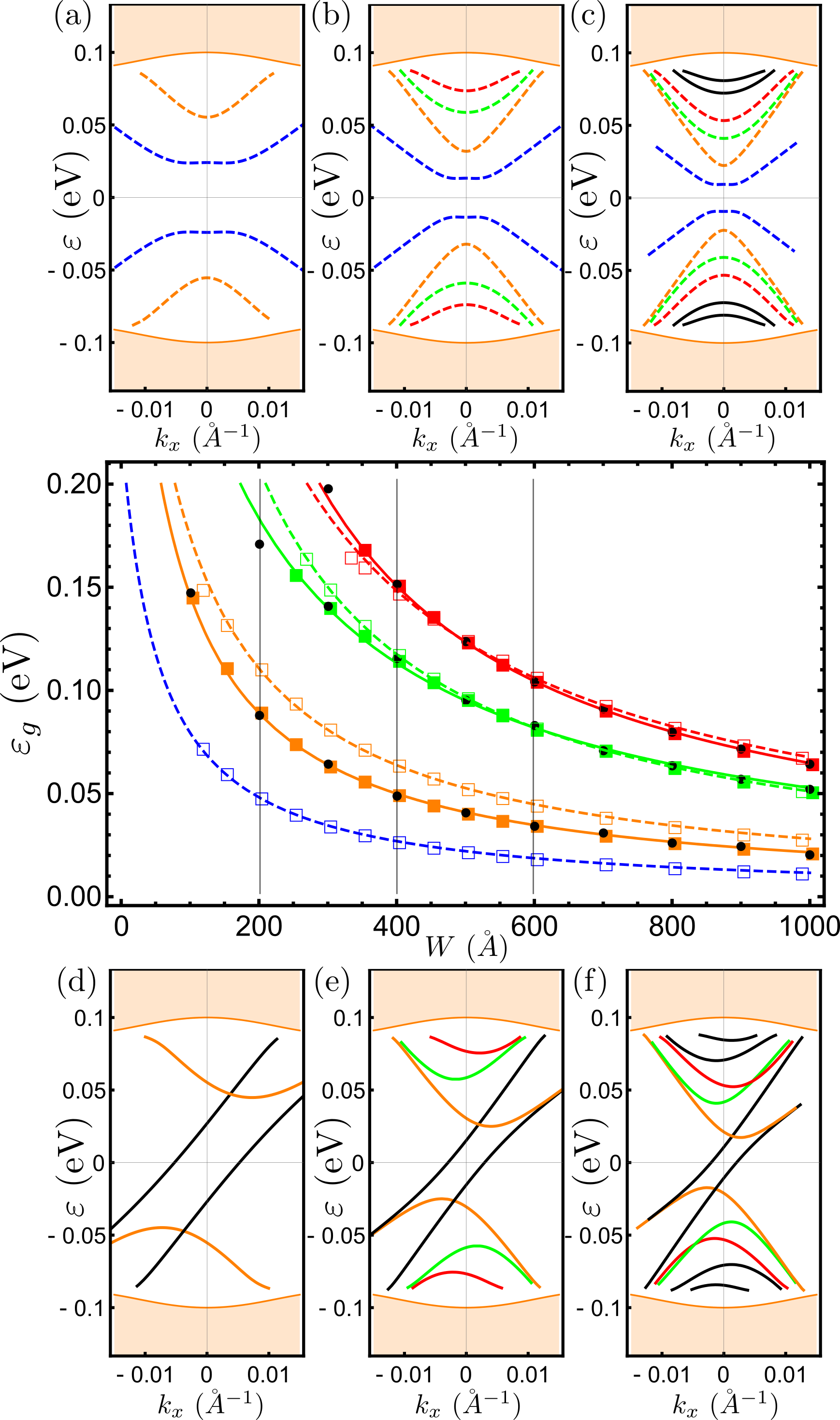}
		\caption{Dependence of energy gaps on delamination width. Empty squares and dashed curves show data and fitting for the AB-2ML-AB system respectively, whilst filled squares and solid curves show data for the AB-2ML-BA system. Black points represent data extracted from the TB model for the AB-2ML-BA system. (a-c) Band structures at W=200\AA, W=400\AA\ and W=600\AA\ respectively for the AB-2ML-AB system. (d-f) Band structures at W=200\AA, W=400\AA\ and W=600\AA\ respectively for the AB-2ML-BA system. Plots are calculated for $\Delta=0.2$~eV and $\delta=U=0$ around the $K^+$ valley.}
		\label{fig09:gaps}
	\end{figure}
	
	Figure \ref{fig09:gaps} illustrates how the energy gaps of the band minima vary with increasing delamination width for both AB-2ML-AB and AB-2ML-BA interlayer stacking, calculated assuming $\delta=0$ for simplicity (though this is not experimentally viable). Fittings are of the form,
	\begin{equation}
		\varepsilon_g=a(W+b)^{-1},
	\end{equation}
	with fitting parameters given in Table.~\ref{paramTable}. Dashed curves are fit to data from systems with the same interlayer stacking in each bilayer (empty squares) whilst solid lines are fit to data from systems with different interlayer stacking to either side of the delamination (filled squares). Black circles show data from the AB-2ML-BA TB model to be in good agreement with the corresponding wavematching model.
	
	The lowest energy (dashed blue) curve corresponds to minima arising from avoided crossings of counter-propagating evanescent states localised at each interface in the AB-2ML-AB system. There is no complementary curve for the AB-2ML-BA system since the localised channels for this configuration are co-propagating and the resulting low-energy bands span the entire gap. The energy gap between higher energy bands follows the same general form, with an especially close agreement between green and red curves in each system.
	
	\begin{table}
		\begin{ruledtabular}
			\begin{tabular}{c c c}
				Curve & a & b \\ \hline
				\textcolor{blue}{- - - -} & 12.2 & 53.6 \\
				\textcolor{myOrange}{- - - -} & 30.1 & 72.3 \\
				\textcolor{green}{- - - -} & 54.2 & 61.2 \\ 
				\textcolor{red}{- - - -} & 74.5 & 102.5 \\
				\textcolor{myOrange}{--------} & 22.7 & 55.9 \\
				\textcolor{green}{--------} & 59.2 & 123.1 \\ 
				\textcolor{red}{--------} & 68.2 & 52.7
			\end{tabular}
		\end{ruledtabular}
		\caption{Fitting parameters for curves in Fig.~\ref{fig09:gaps}.\label{paramTable}}
	\end{table}


\begin{thebibliography}{99}
\bibitem{gamma3Footnote} In contrast to the vertical interlayer hopping, $\gamma_1$, for electrons at the $K^+$ and $K^-$ points in a pristine graphene bilayer the effect of skew interlayer hopping, $\gamma_3$, cancels out as we sum the phase contributions from the three nearest interlayer neighbours. At the delamination such cancellation is broken, which leads to the off-diagonal terms in boundary condition (\ref{matching}). However, the quantitative effect of such a term is negligibly small.
\bibitem{footNote1} Due to the finite nature of the BLG leads in the TB model, additional bands are shown above the bilayer band edge that would not otherwise be present. This leads to a high density of bands along the bilayer band edge, hence the dark region beneath the `mexican hat' dispersion.
\end{thebibliography}
\end{document}